\documentclass[12pt,preprint]{emulateapj}

\def\ApJ{{ApJ}}
\def\MNRAS{{MNRAS}}
\def\AA{{A\&A}}

\def\etal{{et al.~}}

\begin{document}
\title{The bulk Lorentz factor of outflow powering X-ray flare in GRB afterglow}
\author{Zhi-Ping Jin$^{1}$, Yi-Zhong Fan$^{1}$ and Da-Ming Wei$^{1}$}
\affil{Purple Mountain Observatory, Chinese Academy of Sciences, Nanjing 210008, China}
\email{yzfan@pmo.ac.cn(YZF), dmwei@pmo.ac.cn(DMW)}

\altaffiltext{1}{Purple Mountain Observatory, Chinese Academy of Sciences}

\shorttitle{Lorentz factor of X-ray flare}
\shortauthors{Z. P. Jin, Y. Z. Fan \& D. M. Wei}

\begin{abstract}
We develop two methods to estimate the bulk Lorentz factor of X-ray flare outflow. In the first
method the outflow is assumed to be baryonic and is accelerated by the thermal pressure, for which
the final bulk Lorentz factor is limited by the outflow luminosity as well as the initial radius of
the outflow getting accelerated. Such a method may be applied to a considerable fraction of flares.
The second method, based on the curvature effect interpretation of the quick decline of the flare,
can give a tightly constrained estimate of the bulk Lorentz factor but can only be applied to a few
giant flares. The results obtained in these two different ways are consistent with each other. The
obtained bulk Lorentz factor (or just upper limit) of the X-ray flare outflows, ranging from ten to
a few hundred, is generally smaller than that of the Gamma-ray Burst (GRB) outflows.
\end{abstract}


\section{Introduction}
Cosmic Gamma-ray Bursts (GRBs) are the most luminous events ever known in the universe after the Big Bang.
The Luminosity, duration and bulk Lorentz factor are crucial parameters of GRBs \citep{Piran99}. Based on the duration GRBs can be divided into two main groups, short and long bursts. Once the redshift of a burst is measured, with a given  cosmology model the luminosity distance is known, so are its luminosity and isotropic-equivalent $\gamma-$ray energy. Though
the physical processes producing prompt emission and afterglow of GRBs are still unclear, it is widely accepted
that the outflows move ultra-relativistically (i.e., its velocity is very close to the speed of light). The most reliable evidence is the measurement of superluminal movement of the radio afterglow image of GRB 030329 \citep{Taylor04}. In the literature, several
methods have been developed to constrain the initial bulk Lorentz factors of GRB outflows.
(1) The detection of the high energy photons imply the emitting region is optical thin for pair annihilation. This argument leads to a lower limit on Lorentz factor \citep{SP90,LS01}. (2) In some cases the peak in the early GRB afterglow lightcurves marks the deceleration of the external forward shock, at which the Lorentz factor is about half of the initial value. In such cases the initial Lorentz factor, weakly depending on the isotropic-equivalent kinetic energy and the density of the medium, can be tightly constrained
\citep{SP99,Molinari07}. (3) Detection of a thermal
emission component in the GRB afterglow spectrum provides a relatively direct way to estimate the Lorentz
factor. Assuming a thermal radiation efficiency, with the measured temperature and flux of the thermal component, one can calculate the bulk Lorentz factor \citep{Peer07}. (4) The non-detection of a hard X-ray to soft gamma-ray
background emission in GRB prompt stage gives an upper limit on the initial Lorentz factor of GRB
outflows \citep{Zou10}.

Many {\it Swift} GRBs were followed by energetic X-ray flares. Usually their
fluences are only about $1-10\%$ of that of prompt emission. However in quite a few events, the energy of X-ray flare is comparable to that of prompt emission. The temporal behavior and the hardness ratio evolution of X-ray flares are
similar to those of prompt emission pulses \citep{Chincarini10}, supporting the idea that they have the same physical origin of the prompt emission, i.e., they are also due to the activity of central engine \cite[e.g.][]{Fan05,Zhang06}. In such a kind of model, the central engine launches new outflow at late times. As in the prompt emission phase,
the bulk Lorentz factor of the flare outflow is a crucial parameter. Unfortunately, most models constraining the bulk Lorentz factor of the GRB outflow are invalid for the X-ray flares. For example, the methods developed in \citet{SP99} and \citet{Zou10} are irrelevant since there is a preceding and more energetic GRB outflow expanding into the medium.    So far there are three kinds of speculations on the bulk Lorentz factor of the flare outflows: (a) The typical bulk Lorentz factor of flares is just tens and is considerably smaller than that of the outflow powering prompt emission     \citep{Fan05}; (b) The typical bulk Lorentz factor of flares is higher but not much higher than that of the outflow giving rise to prompt emission \citep{Burrows05,Zhang06}. Please note that in both case (a) and case (b), the flare photons are powered by the energy dissipation within the newly launched outflow. (c) In
the X-ray flare model of up-scattered forward shock emission, late outflow with a bulk Lorentz factor $\sim 10^5$ is required \citep{Panaitescu08}. The divergency between these arguments are very large. In this work, we develop two  different methods, as described in section \ref{sec:model}, to estimate the bulk Lorentz factor of the
outflows. The case studies are presented in section \ref{sec:Case}. We summarize our result with some discussion in section \ref{sec:DS}.

\section{The methods}\label{sec:model}

{\bf Method I}: The physical composition of X-ray flare outflows is not well constrained yet.
\citet{Fan05b} suggested that in the neutron star$-$neutron star (or black hole) merger model the
outflow powering X-ray flares following short GRBs might be Poynting-flux dominated since the
fallback accretion onto the newly formed black hole is too low to launch an energetic ejecta via
neutrino annihilation. But for long GRBs the argument is weak. Recently \citet[][and the references
therein]{Chincarini10} argued that the X-ray flare outflow was likely Poynting-flux dominated if
the prompt emission was powered by the magnetic energy dissipation. However the physical origin of
the prompt emission is still in debate. Therefore here we assume that the flare outflow is
baryonic. The central engine releases a large amount of energy into a compact region. Therefore the
outflow is very hot with a temperature $\sim 1~{\rm MeV}$, depending on the total luminosity of the
flare outflow ($L$) as well as the initial radius of the outflow getting launched ($R_0$). As in
the case of fireball powering prompt emission, the flare outflow will be accelerated by the thermal
pressure until it becomes optically thin (i.e., the optical depth $\tau\sim 1$) or saturates at a
radius $R_{\rm f}$, depending on the outflow is baryon-rich or not. In the baryon-poor case the
outflow becomes transparent
 at $R_{\rm f}< \eta R_0$, where $\eta$ is the dimensionless entropy of the initial ejecta. The thermal energy has not been effectively converted into the kinetic energy of the outflow and will give rise to a quasi-thermal radiation component. The absence of such a soft component in the data may suggest that the flare outflow is baryon-rich, for which
the final bulk Lorentz factor can be estimated as $\Gamma_{\rm x}\sim \eta \sim R_{\rm f}/R_{0}$
\citep{Piran99}. On the other hand, the optical depth of the photon at the radius $R_{\rm f}$ can
be estimated as \citep{Paczynski90,Daigne02}:
\begin{equation}
\tau\sim\int^{\infty}_{R_{\rm f}}(1-\beta_{\rm x})n\sigma_{\rm T}dR\sim 1, \label{eq:M1}
\end{equation}
where $n\sim L/4\pi R^2\Gamma_{\rm x} m_{\rm p}c^3$ is the number density of electrons coupled with
protons in the observer's frame, $\sigma_{\rm T}$ is the Thompson cross section, and $m_{\rm p}$ is
the rest mass of protons.
Combing with the relations $R_{\rm f}\sim \Gamma_{\rm x} R_{0}$ and $\beta_{\rm
x}\sim1-1/2\Gamma_{\rm x}^{2}$, eq.(\ref{eq:M1}) gives
\begin{equation}
\frac{L\sigma_{\rm T}}{8\pi\Gamma_{\rm x}^{4}m_{\rm p}c^{3}R_{0}}\sim 1. \label{eq:M2}
\end{equation}
Therefore the bulk Lorentz factor of X-ray flares is related to $L$ and $R_{0}$ as
\citep[e.g.][]{Meszaros00,Nakar05,Fan10}
\begin{equation}
\Gamma_{\rm x} \leq \Gamma_{\rm max}=5\times10^{2} L_{50}^{1/4}R_{0,6}^{-1/4},
\label{eq:Gamma_x}
\end{equation}
throughout this work, the convention $Q_{\rm x}=Q/10^{x}$ has been taken into account except for
some special notations. For $\eta \ll \Gamma_{\rm max}$, the photospheric radius of the flare
outflow is estimated to be $\sim 6\times 10^{10}~{\rm cm}~L_{50}\eta_{2}^{-3}$
\citep{Paczynski90,Fan10}, much larger than $R_{\rm f} \sim 10^{8}~{\rm cm}~\eta_{2}R_{0,6}$. With
a proper $R_0$ and the observed X-ray flare luminosity $L_{\rm x}$ ($\sim \epsilon_{\rm x}L$, where
$\epsilon_{\rm x}$ is the X-ray flare efficiency and is taken to be comparable to that of GRBs,
i.e., $\sim 0.1$ \citep[e.g.][]{Fan06}), we are then able to give an upper limit on $\Gamma_{\rm
x}$. The advantage of this method is that it may apply to a good fraction of X-ray flares. The
limit of this approach is that it is only valid for the baryonic outflows.

{\bf Method II}: The quick decline of the X-ray flares \citep{Piro05,Burrows05} may have imposed a
tight constraint on the emission radius $R_{\rm x}$ \citep{Zhang06,Lazz06,DaiX07}, provided that
the quickly decaying X-ray emission is the high latitude component of the flare pulses \citep[][in
some literature it is called the ``curvature effect"]{Fenimore96,KP00}. This interpretation, if
correct, requires a very large variability timescale $\delta T$ \citep[][see also below]{Fan05}.
The corresponding bulk Lorentz factor thus can be estimated by $\Gamma_{\rm x} \approx [R_{\rm
x}/(2 c \delta T)]^{1/2}$, please note that the timescales involved here and below are all in the rest
frame of the GRB source.
If $R_{\rm x}$ as well as $\delta T$ can be reliably estimated, $\Gamma_{\rm x}$ can be reasonably constrained.

For simplicity, we take the leading late internal shock model for
illustration. Assuming that there are two shells, the slow one is
ejected at $T$ and moves with a bulk Lorentz factor $\Gamma_{\rm s}$,
the faster one is with $\Gamma_{\rm f}$, and their ejection interval is $\delta t$ \citep[see also][]{KPS97}. The
widths of both shells are $\sim \Delta$ (in the observer's frame). The faster one would catch up with the slower at a radius
\begin{equation}
R_{\rm coll} \approx 2\Gamma_{\rm s}^2 c \delta t.
\end{equation}
After the merger, the newly formed shell has a bulk Lorentz factor $\Gamma_{\rm x} \sim
\sqrt{\Gamma_{\rm f} \Gamma_{\rm s}}$, if the mass of these two shells are comparable
\citep{Piran99}. The timescale of that merger is determined by the reverse shock crossing the fast
shell and is $\sim \Delta/c$. The corresponding radius is
\begin{equation}
R_{\rm x} \sim (2\Gamma_{\rm s}^2 c \delta t +2\Gamma_{\rm x}^2 \Delta),
\end{equation}
at which the emission peaks. In the internal shock phase, usually
the electrons cool rapidly \citep{Piran99}. After the ceasing of the
internal shock, the emission is from the high latitude
$\theta>1/\Gamma_{\rm x}$ and takes a form \citep{Fenimore96,KP00,Fan05}
\begin{equation}
F\propto (T/\delta T)^{-(2+\beta)}.
\label{eq:Fenimore}
\end{equation}
where $\delta T \approx R_{\rm x}/(2\Gamma_{\rm x}^2 c) \approx \Delta/c$ for $\Gamma_{\rm x}^2 \gg
\Gamma_{\rm s}^2$ and $\delta t \sim \Delta/c$, {and the standard convention of the flux as a
function of both time and frequency $f_{\nu}\propto T^{-\alpha}\nu^{-\beta}$ is adopted}. In this
work, we assume that the ejecta is uniform.

In reality, the flare consists of many (for example $k$) pulses. After the ceasing of internal shocks
at $T_{\rm t}$, what we observe is the high latitude
emission of the early pulses.  The X-ray flux declines as \footnote{Please bear in mind that the
following discussion is valid as long as the prompt emission consists of many pulses. These pulses
could be powered by either late internal shocks or late internal magnetic energy dissipation. So the validity of our method II, different from method I,
is independent of the physical composition of the outflow.}
\begin{equation}
F_{\rm X} \approx {\sum_{\rm i=1}^{k}} F_{\nu_{\rm x},i}
[(T-T_{0,i})/\delta T_i]^{\rm -(2+\beta_i)},
\end{equation}
where $i$ represents the $i-$th pulse, $F_{\nu_{\rm x},i}$ is the peak emission of the $i-$th
pulse, $T_{0,i}$ is the time when the $i-$th pulse is ejected. Such a decline is much steeper than
$(T /T_{\rm t})^{-(2+\beta)}$ as long as $T_{\rm t}\gg \max \{\delta T_i \}$ (see Fig.2 of
\citet{Fan05} for illustration). {\it This can be understood as follows}. In the popular internal
shock scenario, each prompt pulse is independent and is emitted at a radius $\sim R_{\rm x}$. At
late times, the emission contributed by early shells is from a very large angle, and is very weak
due to the relativistic beam effect. {\it The observed flux is thus dominated by the curvature
emission component of the last pulse (if all $\delta T_i$ are comparable) or one early pulse having
a very large duration $\sim T_{\rm t}$.} As a result, the net flux of these shells can be
approximated by
\begin{equation}
F_{\rm X} \approx F_{\nu_{\rm x},k} [(T-T_{0,k})/\delta T_{\rm k}]^{\rm
-(2+\beta_k)}, \label{eq:FP2}
\end{equation}
which could be far more sharply than $(T/T_{\rm
t})^{-(2+\beta)}$ since the internal shocks cease.

As shown in \citet{Liang06} and \citet{zbb07}, the sharp decline of some X-ray flares are well fitted by $F_{\rm x} \propto (T-T_0)^{-(2+\beta)}$, which is consistent with
the curvature effect \citep{Fenimore96,KP00} interpretation of that phenomenon. The inferred $T_0$ is a good approximation of the ejection time of the last
dominant pulse ({rather than the re-activity time of the central
engine unless there is just a pulse}).

As shown in eq.(\ref{eq:FP2}), for the last dominant pulse, $\delta T \sim T_{\rm p}-T_0$, where $T_{\rm p}$
is the peak time of the last dominated pulse. On the other hand, the curvature emission component
(i.e., the tail emission) lasts $T_{\rm tail}$ until the edge of the ejecta is visible
(see Fig.\ref{fig:tail}). For an ejecta having a half-opening angle $\theta_{\rm j}$ (which can be
estimated from the achromatic breaks of the late afterglow lightcurves \citep{Rhoads99}), the prompt
emission radius can be estimated as
\begin{equation}
R_{\rm x} \approx {cT_{\rm tail} \over (1-\cos \theta_{\rm j})}.
\end{equation}
So we have
\begin{equation}
\Gamma_{\rm x} \approx [{T_{\rm tail} \over   2(1-\cos \theta_{\rm j})\delta
T}]^{1/2}\approx [{T_{\rm tail} \over   2(1-\cos
\theta_{\rm j})(T_{\rm p}-T_0)}]^{1/2}. \label{eq:FP3}
\end{equation}
In many cases, it is inconvenient to get $T_{\rm p}$ and $T_0$, respectively. Fortunately,  a simpler way to estimate
$T_{\rm tail}/\delta T$ is available. We introduce the factor ${\cal R}$ to denote the ratio
between the peak flux of the X-ray flare ($F_{\rm x,p}$) and the flux when a cutoff emerges
($F_{\rm x,c}$), i.e., ${\cal R}\equiv F_{\rm x,p}/F_{\rm x,c}$ (see Fig.\ref{fig:tail}). With
eq.(\ref{eq:FP2}), we have
\begin{equation}
{\cal R} \approx ({T_{\rm tail} \over \delta T})^{2+\beta}
\Rightarrow {T_{\rm tail} \over \delta T} \approx {\cal R}^{1\over
2+\beta}. \label{eq:FP4}
\end{equation}
Combing eq.(\ref{eq:FP3}) with eq.(\ref{eq:FP4}), we have
\begin{equation}
\Gamma_{\rm x} \approx {\cal R}^{1/[2(2+\beta)]} / \theta_{\rm j}.
\label{eq:FPB}
\end{equation}
With a typical $\beta \sim 1$, $\Gamma_{\rm x} \propto {\cal R}^{1/6}$. So our estimate of
$\Gamma_{\rm x}$ can not be modified significantly by the uncertainty of ${\cal R}$. The choice of
$\theta_{\rm j}$, instead, is crucial for our purpose. For the X-ray flare outflow, there is no
simple/reliable method to estimate its half-opening angle $\theta_{\rm j, flare}$. However, it is
reasonable to argue that $\theta_{\rm j, flare}$ can not be much smaller than $\theta_{\rm j,
GRB}$. This is because X-ray flare(s) have been observed in about $40\%$ {\it Swift} GRBs, so
statistically speaking, $\theta_{\rm j, flare}\sim \sqrt{0.5}\theta_{\rm j, GRB}\sim 0.7\theta_{\rm
j, GRB}$ (X. Y. Dai 2006, private communication), where $\theta_{\rm j,GRB}$ is the half-opening
angle of the GRB ejecta that can be inferred from the late afterglow break. Such a small correction
can just increase our estimate of $\Gamma_{\rm x}$ (see eq.(\ref{eq:FPB}) for details) by a factor
of 1.4 and thus does not change our basic conclusion. As a consequence, in the case studies (see
below), we simply take $\theta_{\rm j}\equiv \theta_{\rm j, flare}=\theta_{\rm j,GRB}$. The error
of the resulting $\Gamma_{\rm x}$ can be estimated straightforwardly and can be approximated by
$\delta \Gamma_{\rm x} \sim \Gamma_{\rm x}\sqrt{{1\over 4(2+\beta)^{2}}({\delta {\cal R} \over
{\cal R}})^{2}+({\delta \theta_{\rm j}\over \theta_{\rm j}})^{2}}$, where $\delta {\cal R}$ and
${\delta \theta_{\rm j}}$ are the errors of ${\cal R}$ and $\theta_{\rm j}$, respectively. For
$\beta \sim 1$ and $\delta {\cal R}/{\cal R} \leq 6 \delta \theta_{\rm j}/\theta_{\rm j}$, we have
$\delta \Gamma_{\rm x} \sim \Gamma_{\rm x} \delta \theta_{\rm j}/\theta_{\rm j}$. For a
conservative estimate $\delta \theta_{\rm j}/\theta_{\rm j} \sim 1/2$, we have $\delta \Gamma_{\rm
x} \sim 0.5\Gamma_{\rm x}$.

Different from method I, the current method can give a somewhat reliable estimate of $\Gamma_{\rm x}$
rather than an upper limit. However, a cutoff at the end of the quick decline is needed to achieve that goal.
Unfortunately an unambiguous cutoff in the quick decline phase has just been identified in a few giant flares, limiting the application of method II. We'd like to point out that method II is also valid to estimate the bulk Lorentz factor
of the late outflow of GRBs if the quick decline phase of prompt emission is also attributed to the so-called curvature effect.

\section{Case studies}\label{sec:Case}

\begin{figure}
\begin{center}
\includegraphics[width=0.5\textwidth, clip=true]{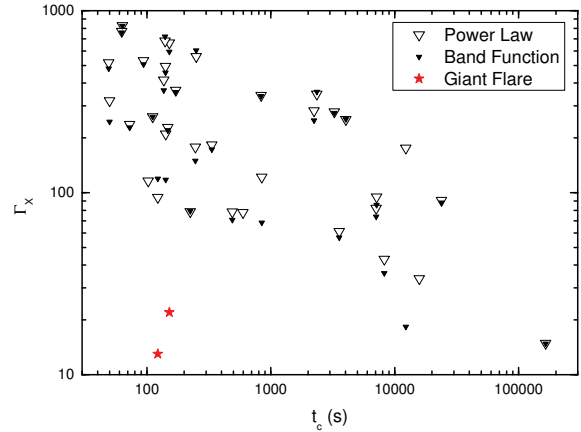}
\caption{Upper limits of the bulk Lorentz factor of outflow powering X-ray flare in GRB afterglow
and the center time of the X-ray flares. The two types of triangles represent the upper limits on
the bulk Lorentz factor estimated in method I. The hallow and filled triangles are for the single
power-law spectrum assumption and the Band function spectrum assumption, respectively. Stars are
the bulk Lorentz factor, estimated in method II, of two giant flares.} \label{fig:gamma_t}
\end{center}
\end{figure}

\begin{figure}
\begin{center}
\includegraphics[width=0.5\textwidth, clip=true]{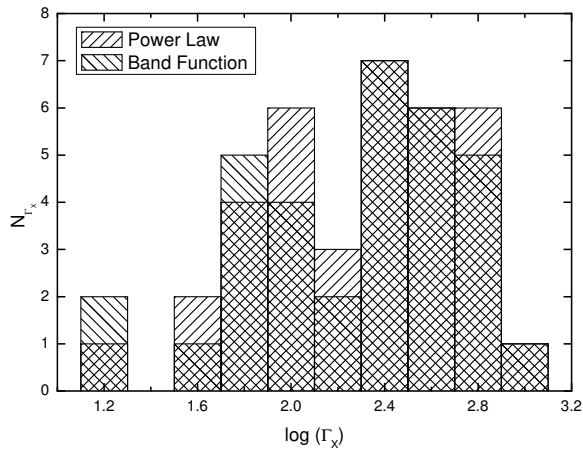}
\caption{The distribution of the upper limits of bulk Lorentz factor of outflow powering X-ray
flare in GRB afterglow. Two types of slant lines are for the single power-law spectrum assumption
and the Band function spectrum assumption, respectively.} \label{fig:distribution}
\end{center}
\end{figure}

For method I, we take a sample consisting of 36 X-ray flares detected in 14 GRBs with measured
redshift, as reported in \citet{Chincarini07} and \citet{Falcone07}. The average bolometric
luminosity, i.e., the bolometric energy divided by the duration, is derived for both the power-law
and the Band function fits to the flares \citep{Falcone07}. As already mentioned, the total
luminosity of the flare is taken to be 10 times than that of the X-ray emission. $R_{0}$ is taken
as $10^{7}$ cm, which is comparable to the radius of a neutron star or the last stable circle orbit
for a rapidly rotating black hole (In the thermal radiation modeling of some GRBs, $R_0$ is
estimated to be $\sim 10^{8}$ cm or even larger \citep[e.g.][]{Peer07}. So our estimate is likely
conservative). The results are presented in Fig.\ref{fig:gamma_t} and Fig.\ref{fig:distribution}.
The resulting upper limits on $\Gamma_{\rm x}$ are between tens to hundreds, which are generally
smaller than the initial bulk Lorentz factor of GRB outflows $\sim 10^{2}-10^{3}$
\citep[e.g.][]{Piran99,LS01,Molinari07,Zou10}. For the baryonic outflow, we have $\Gamma_{\rm x}
\sim L/\dot{M} c^{2}$, where $\dot{M} \sim 5\times 10^{-6}L_{50}~\Gamma_{\rm x,1}^{-1}M_\odot~{\rm
s}^{-1}$ is the mass loading rate of the outflow. The inferred $\Gamma_{\rm x}\geq 10$ following
the flares in GRB 050502B and GRB 050724 suggests that the flare outflows are still relativistic,
strengthening the connection between the X-ray flares and the prompt soft $\gamma-$ray emission.
Since the typical total luminosity of the flare outflows is about two or more orders of magnitude
lower than that of the GRB outflows, the mass loading rate of the flare outflow is expected to be
lower than that of the GRB outflow otherwise relativistic ejecta can not be launched. This is
reasonable since the pole region of a dying massive star (or the remnant of the merger of two
compact objects) is expected to be more and more clean as the time goes on.

Below we focus on method II. The sharp decline has been well detected in a good fraction of X-ray flares. For
our purpose, a cutoff, emerging when the emission at the edge of the ejecta is within the line of sight,
of the tail emission is needed to constrain $R_{\rm flare}$ reliably. Such a cutoff has just been possibly identified in a few cases, for example, the giant
flare in GRB 050502B \citep{Falcone06}, and the giant flare in GRB 050724 \citep{Camp06}. The reason
for such a rare detection is the following. In reality, the X-ray emission is not only contributed
by the high latitude emission but also contributed by the forward shock emission. The latter would
dominate over the former when the high latitude emission component has dropped by 2 or
more orders (correspondingly, $T_{\rm tail} \sim 5 \delta T$ for $\beta\sim 1$) unless the X-ray
flare is strong enough, or the forward shock emission is very dim. On the other hand, when
the flux has dropped to a level of $\sim 10^{-11}~{\rm erg~s^{-1}~cm^{-2}}$, the signal/noise
ratio is not high enough to get a good quality detection and thus renders the identification of the
cutoff difficultly. The data (together with corresponding reference) are presented in the
Table.\ref{tab:Tab1}.

\begin{table*}
\caption{The physical parameters, including the estimated bulk Lorentz factor,  of giant flares
detected in GRB 050502b and GRB 050724.}
\begin{tabular}{cccccccccc}
\hline GRB & $t_{\rm c}$ & $\beta$ & ${\cal R}$ &
 $\theta_{\rm j}$ & $\Gamma_{\rm x}$(I) & $\Gamma_{\rm x}$(II) & Reference\\
\hline
050502b & 152s & 1.3 & 1100 &  $\sim 0.13$ & $<666$ & 22 & \citet{Falcone06}\\
050724 & 122s & 1.34 & 72 &  $\sim 0.15$ & $<120$ & $13$ & \citet{Camp06,Pana06}\\
\hline
\end{tabular}\label{tab:Tab1}
\end{table*}

{\it Note} that there is a possible independent argument favoring a $\Gamma_{\rm x} \sim$ tens for
giant flares following GRB 050502b and GRB 050724. As suggested in \citet{Fan05}, in the leading
late internal shock model, the X-ray flare outflow with a $\Gamma_{\rm x} \sim $ tens will catch up
with the initial GRB outflow when the latter has swept a large amount of material and then got
decelerated. Such an energy injection process would give rise to a flattening (if there is also a
wide range of the bulk Lorentz factors of the X-ray flare outflow) or re-brightening signature (if
the range of the bulk Lorentz factors of the flare outflow is narrow). The re-brightening has been
well detected in the late X-ray afterglow of GRB 050502b and GRB 050724. To attribute the very late
X-ray flare in GRB 050502b to an energy injection caused by the giant X-ray flare outflow, a
$\Gamma_{\rm x} \sim 20$ is needed \citep{Falcone06}, matching our result perfectly. As for GRB
050724, \citet{Pana06} showed that the late multi-wavelength re-brightening could be well
reproduced by an energy injection. Before the energy injection, the kinetic energy of the GRB
ejecta $E_{\rm k} \sim 10^{50}$ erg and the number density of the medium is $n \sim 0.1~{\rm
cm^{-3}}$. The decelerating GRB ejecta has a bulk Lorentz factor $\sim 10~(E_{\rm
k,50}/n_{-1})^{1/8}(T/10^4{\rm s})^{-3/8}$. If this energy injection is also caused by the flare
outflow catching up with the GRB ejecta, a $\Gamma_{\rm x} \sim 10$ is needed, which is consistent
with our result.

\begin{figure}
\begin{center}
\includegraphics[width=0.5\textwidth, clip=true]{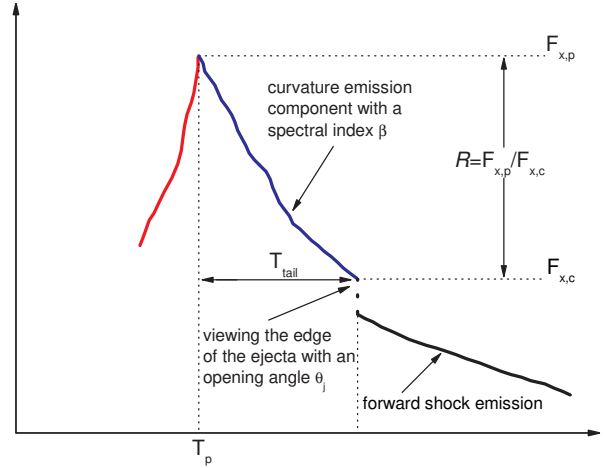}
\caption[...]{A schematic plot of the curvature emission component (tail emission) that is useable
for method II. } \label{fig:tail}
\end{center}
\end{figure}

Finally, we would like to
point out that only the X-ray flares brighter than the forward shock X-ray emission can be
identified. There could be some faint X-ray flares which have even lower bulk Lorentz factor.

\section{Discussion and Summary}\label{sec:DS}
Bright X-ray flares have been well detected in the afterglow of a considerable fraction of
GRBs. The radiation mechanism powering these events is still unclear. A widely accepted
hypothesis is that new outflow should be launched by the central engine and the flares should be
powered by the energy dissipation within the new outflow. The lack of an optical flare associated with
 the X-ray flare is in general consistent with such a scenario \citep{Fan05}. The bulk Lorentz factor
of the newly launched outflow is a crucial parameter for us to understand X-ray flares. For
example, with the inferred bulk lorentz factor we can see whether the outflow is relativistic or
not. Then we can estimate the corresponding mass loading rate in the pole region of the central
engine. If the inferred bulk Lorentz factor is so high that surpasses the upper limit given in
eq.(\ref{eq:Gamma_x}), the baryonic outflow model will be ruled out.

In this work we have developed two methods to estimate the bulk Lorentz factor of X-ray flare
outflow (see section 2). In the first method the outflow is assumed to be baryonic and is
accelerated by the thermal pressure, for which the final bulk Lorentz factor is limited by the
outflow luminosity as well as the initial radius of the outflow getting accelerated. Such a method
may be applied to a considerable fraction of flares. However it can only give an upper limit and is
invalid if the flare outflow is Poynting-flux dominated. The second method, based on the curvature
effect interpretation of the quick decline of flare, is independent of the physical composition of
the outflow and can give a better constrained estimate of the bulk Lorentz factor of the flare
outflow but can only be applied to a few events \footnote{There is the other method without the
need of the nature of the outflow to constrain the bulk Lorentz factor. The idea is the following.
In the synchrotron radiation model, the typical synchrotron radiation frequency, the cooling
frequency as well as the synchrotron self-absorption frequency are functions of the radius of the
flare emission and the bulk Lorentz factor. Therefore with the measured redshift and the spectrum
in a very wide energy range in principle we can constrain the bulk Lorentz factor. A reliable
infrared/optical to X-ray spectrum of the flare, however, is hard to obtain. For example the
infrared/optical emission of the flare is found to be usually outshone by simultaneous forward
shock emission.}. The obtained bulk Lorentz factors (or just upper limit) of the X-ray flare
outflows in these two different ways are consistent with each other and are generally smaller than
that of the GRB outflows (see section 3). However, the flare outflows are still relativistic and
the corresponding baryon loading rate is expected to be very low ($\dot{M} \lesssim
10^{-5}~M_\odot~{\rm s}^{-1}$). This finding, together with other results from the analysis of the
X-ray flare data such as the temporal behavior and the hardness ratio evolution, strongly favors
the hypothesis that X-ray flares are the low energy analogy of the prompt soft $\gamma-$ray
emission \citep{Fan05,Zhang06,Chincarini10,Margutti10,Shao10}. With method II, the Lorentz factors
of the two giant flares are found to be $\sim {\rm a ~few}\times 10$, well below the upper limits
given by method I (see also Tab.1). Hence the baryonic outflow model is not challenged. For the
magnetic outflow with reasonable baryon loading, a moderate Lorentz factor is also possible.
Further data like the high linear-polarization of the X-ray photons is needed to claim the magnetic
nature of the flare outflow.

If the X-ray flare outflows do have a bulk Lorentz factor $\sim 10^{5}$, as suggested in the flare
model of up-scattered forward shock emission \citep{Panaitescu08}, the acceleration can not be due
to the thermal pressure. The only known such a kind of extreme astrophysical object is the pulsar
wind, which may move with a Lorentz factor as high as $\sim 10^{6}$ and its acceleration is likely
a result of the significant magnetic energy reconnection \citep{Kennel84}. It's less likely to be
the case for the X-ray flares since the central engine of GRBs is not in a cavity as clean as that
of a pulsar at an age of $\geq 10^{3}$ years. As for the flare model of up-scattered forward shock
emission, one more challenge is that the similarities between the GRB prompt emission and the X-ray
flare emission strongly suggest a similar physical origin of these two kinds of phenomena.

\section*{Acknowledgments}
We thank X. Y. Dai for helpful communication. This work was in part supported by the National
Natural Science Foundation (grants 10673034 and 10621303) of China, and by the National 973 Project
on Fundamental Researches of China (2007CB815404 and 2009CB824800).

\end{document}